# Effect Size-Driven Pathway Meta-Analysis for Gene Expression Data

Juan Antonio Villatoro-García[a,b,*], Pablo Pedro Jurado-Bascón[a,b,*], Pedro Carmona-Sáez [a,b]

[a] *Department of Statistics and Operations Research, University of Granada, Granada, Spain.*
[b] *GENYO. Centre for Genomics and Oncological Research: Pfizer / University of Granada / Andalusian Regional Government, PTS Granada, 18016, Granada, Spain.*
*\* Joint First Authors*

JAVG: juan.villatoro@genyo.es
PPJB: pjurado@ugr.es
PCS: pcarmona@ugr.es

Correspondence
Pedro Carmona-Sáez, PhD. *Department of Statistics and Operations Research, University of Granada* Faculty of Medicine. Av. Doctor Jesús Candel Fábregas nº 11 18071 Granada

Email: pcarmona@ugr.es

Data availability

Gene expression data is publicly available at GEO repository. GSEMA R package is available at CRAN and GitHub repository.

Funding

This work has been funded by grant PID2020-119032RB-I00 funded by MCIN/AEI/10.13039/501100011033. JAVG was funded by the Teaching Staff Program, Ministerio de Universidades (grant number FPU19/01999).

Conflict of interest

The authors declare no conflict of interest.

# Abstract

The proliferation of omics datasets in public repositories has created unprecedented opportunities for biomedical research but has also posed significant challenges for their integration, particularly due to missing genes and platform-specific discrepancies. Traditional gene expression meta-analysis often focuses on individual genes, leading to data loss and limited biological insights when there are missing genes across different studies. To address these limitations, we propose GSEMA (Gene Set Enrichment Meta-Analysis), a novel methodology that leverages single-sample enrichment scoring to aggregate gene expression data into pathway-level matrices. By applying meta-analysis techniques to enrichment scores, GSEMA preserves the magnitude and directionality of effects, enabling the definition of pathway activity across datasets. Using simulated data and case studies on Systemic Lupus Erythematosus (SLE) and Parkinson's Disease (PD), we demonstrate that GSEMA outperforms other methods in controlling false positive rates while providing meaningful biological interpretations. GSEMA methodology is implemented as an R package available on CRAN repository.

# 1.Introduction

High-throughput omics technologies have revolutionized our understanding of biological systems by enabling the systematic quantification of variables such as genes, transcripts or proteins in organisms and individual cells. Over the past two decades, the increasing accessibility and adoption of these technologies have resulted in a rapid expansion of available omics datasets stored in public repositories[1] such as the National Center for Biotechnology Information (NCBI) Gene Expression Omnibus (GEO)[2] or ArrayExpress[3].

This amount of data has sparked growing interest in the development of advanced data integration methods to derive new scientific insights. Among these, meta-analysis techniques have been an important focus of research in recent years and have been widely used to jointly analyze multiple datasets to derive a common and significant outcome. In the analysis of gene expression data, meta-analysis enables the identification of shared molecular signatures across datasets, enhancing the reproducibility of findings and facilitating the discovery of robust biomarkers[4].

Typically, gene expression meta-analysis involves computing a common effect for individual genes across different studies to identify those genes that show a consistent differential expression pattern. The outcome is therefore a prioritized list of genes that can be subsequently analyzed to find enriched functional annotations with methods such as Gene Set Enrichment Analysis techniques (GSEA)[5]. These methods identify biological pathways (or gene sets) linked to the gene list, offering valuable insights into the underlying biological processes. However, this traditional approach[6] to gene expression meta-analysis might face significant challenges. A primary limitation stems from the issue of missing data, where genes present in one study may be absent in others[7]. The most extended approach is to discard from the meta-analysis genes that are not present in all platforms, which often results in a substantial loss of information, potentially overlooking critical biological insights[7]. To address this limitation, various solutions have been proposed to manage missing genes, primarily focusing on the imputation of unmeasured genes[7–9]. While these techniques have demonstrated good performance[8,9] their results may show an inherent bias when a large number of values are imputed[8]. Additionally, the integration of datasets generated on different platforms, such as RNA sequencing (RNA-seq) and microarrays, might introduce technical biases as expression values from these platforms are not directly comparable, complicating their integration without additional adjustments.

An alternative approach for traditional gene expression meta-analysis that can overcome these limitations is to integrate results of pathway enrichment analyses rather than gene expression data. That is, to perform a meta-analysis on the gene set space rather than the gene space. Different methodologies have been proposed to this end. Meta-Analysis of Pathway Enrichment (MAPE) involves applying a meta-analysis based on the combination of p-values in genes or pathways from independent studies to obtain the pathway enrichment results[10]. Chen et al.[11] developed a methodology that allowed combining information from gene sets and expression data through Bayesian modeling[11]. Lu et al. implemented iGSEA, which calculates its own enrichment score as an effect and uses it for effect estimation in an adaptive effects model that allows for both fixed and random effects[12].

Despite their strengths, these methods often integrate data at p-value level, which discard information about effect sizes and their directionality, such as activation or inhibition of biological pathways. This limitation hampers the ability to fully capture the biological relevance of the data, particularly when comparing results across diverse studies or platforms.

In this work we propose an alternative strategy based on the application of single-sample enrichment (SSE) scoring schema, which calculates pathway activity scores at the level of individual samples. This approach enables the aggregation of gene expression data into pathway-level matrices, where each row represents a pathway rather than a gene. By working with pathway activity scores instead of raw gene expression values, this methodology alleviates the challenges associated with missing data and platform-specific discrepancies. Furthermore, pathway activity scores are directly comparable across datasets, facilitating the identification of differentially expressed pathways with greater consistency and accuracy.

This new methodology, termed GSEMA (Gene Set Enrichment Meta-Analysis) leverages the strengths of meta-analysis techniques for combining effect sizes but applies them to enrichment scores derived from pathway matrices. Unlike traditional methods, GSEMA preserves both the magnitude and directionality of effects, ensuring that the biological interpretation of pathway activity—whether activation or inhibition—is retained. By working at the pathway level, GSEMA minimizes the impact of missing genes and enhances the biological relevance of the results. Additionally, the approach allows for seamless integration of datasets from diverse transcriptome platforms by replacing raw expression values with pathway enrichment scores.

We demonstrate the performance of the GSEMA methodology using simulated data and two case studies: Systemic Lupus Erythematosus (SLE) and Parkinson's Disease (PD). In these analyses, GSEMA performed well in controlling the false positive rate while yielding significant biological insights compared to other methods. The GSEMA methodology is implemented in the R package which is available on CRAN: https://cran.r-project.org/web/packages/GSEMA/index.html and on GitHub: https://github.com/Juananvg/GSEMA.

## 2. Methods

### 2.1. Calculation pathway activity at a single sample level

Let start with *K* studies each with an gene expression matrix, $M_{G \times N}$, where *G* is the total number of genes, *N* is the total number of samples (patients) and $m_{ij}$ is the expression of the *i-th* gene in the *j-th* sample. To perform the gene set meta-analysis, we transform the different expression levels into gene matrices, $M_{P \times N}$, where *P* is the total number of gene sets. Different single sample enrichment (SSE) methodologies have been developed to obtain an enrichment score per gene set for each sample from the gene expression matrix. We have applied four different techniques: single sample Gene Set Enrichment analysis (ssGSEA)[13], Gene Set Variation Analysis (GSVA)[14], Zscore Gene Set Enrichment Analysis[15] (called Zscore from now on) and singscore[16].

Single sample Gene Set Enrichment (ssGSEA)[13]: Absolute expression values $\left(\left|m_{ij}\right|\right)$ of a given patient (column of $M_{G \times N}$) are rank-normalized in decreasing order and stored in a list *L*. For that patient and a particular gene set, *L* is divided in two groups, those outside and inside of the gene set. Then, the Empirical Cumulative Distribution Function (ECDF) is computed for each group. The ECDF for the first group is calculated using the standard form, while the ECDF for the second group is weighted by its values in L. The enrichment score for that patient and gene set is the sum of the differences between the two group´s ECDFs.

Zscore[15]: The $m_{ij}$ expression values are standardized by rows (genes), giving $z_{ij}$. For each patient, the different $z_{ij}$ values are combined using a methodology similar to Stouffer's method applied in the combination of p-values[17,18]. To achieve this, the different $z_{ij}$ values are summed, and subsequently, the sum is divided by the square root of the number of genes that compose the gene set. Thus, each patient obtains a z-score value as enrichment score for each gene set.

Gene Set Variation Analysis (GSVA)[14] : Considering each row of $M_{G \times N}$ as the expression profile of its gene, we calculate its cumulative distribution function via gaussian or Poisson kernel estimation ($\hat{F}_i$), for microarray and RNA-Seq respectively, and reassign each value of $M_{G \times N}$ ($m_{ij}$) with that of the estimated function ($\hat{F}_i(m_{ij}) = z_{ij}$). The normalization step takes those $z_{ij}$ values ranked by rows, centers them and apply absolute value. These normalized values are then used, per patient, for the Kolmogorov-Smirnov like statistic of the original GSEA[5] given a gene set, resulting in the enrichment score.

Singscore[16]: as in ssGSEA, the expression values of a patient are ranked, but instead of using the absolute value, singscore considers the direction of the expected effect (increasing or decreasing for up-regulated and down-regulated gene sets respectively). For a given gene set, the mean of the ranks belonging to that set is calculated and normalized with the median and the theoretical minimum and maximum of the mean rank, that is repeated for every patient and gene set. The mean rank is calculated differently if the direction of the effect is not known beforehand in order to mark as relevant the absolute deviation from the median.

## 2.2. Meta-analysis based on the effects size combination

### 2.2.1. Calculation of the effect size

Once our expression matrices have been transformed to Gene Set Enrichment Matrices, we can now apply techniques of gene expression meta-analysis based on the combination of effect sizes to integrate the different datasets. The choice of effect size calculation depends on the nature of the data under consideration[4]. In our case, as occurs in the gene expression meta-analysis, we employ the standardized mean difference, known as *Hedges' g*, which was obtained from *Cohen's d*, both are defined as[19]:

$$d_{ij} = \frac{\bar{X}_{ijE} - \bar{X}_{ijC}}{S_{ij}} \quad (1)$$

$$g_{ij} = J_j \times d_{ij} \quad (2)$$

Where:

- $d_{ij}$ is *Cohen's d* for the pathway *i* in the study *j*.
- $g_{ij}$ is *Hedges' g* for the pathway *i* in the study *j*.
- $\bar{X}_{ijE}$ and $\bar{X}_{ijC}$ are the pathway *i* mean score values for the experimental and control group, respectively in the study *j*.

- $S_{ij} = \sqrt{\frac{(n_{jE}-1)S^2_{ijE}+(n_{jC}-1)S^2_{ijC}}{n_{jE}+n_{jC}-2}}$, where $n_{jE}$, $n_{jC}$ and $S^2_{ijE}$ and $S^2_{ijC}$ are the sample sizes and variances of the experimental and control group respectively for the pathway $i$ in the study $j$.
- $J_j = 1 - \frac{3}{4 \times (n_{jE}+n_{jC}-2)-1}$, is a correction factor for a known theoretical *Cohen's d* bias in the study $j$.

The variance of the Hedges' g estimator is:

$$V_{g_{ij}} = J_j^2 \times \left(\frac{n_{jE}+n_{jC}}{n_{jE} \times n_{jC}} + \frac{d_{ij}^2}{2(n_{jE}+n_{jC})}\right) \quad (3)$$

Although this is the most common measure in gene expression meta-analysis, in the context of the differential expression analysis, numerous methods have been developed to allow for a more accurate and appropriate analysis of this difference in means. Specifically, one of the most widely used methods is the *moderated t-test* [20] included in the widely used R package *limma*[21,22]. The *moderated t-test statistic*, unlike the traditional *t-test statistic*, calculates the variance using information from the rest of the variables (genes) in such a way that it corrects for false positives due to small differences with small variances. On the other hand, *Rosenthal and Rosnow*[23] demonstrate that from a *t statistic* and its degrees of freedom, the corresponding estimator of Cohen's d can be calculated using the following expression:

$$d_{ij} = \frac{(n_{jE}+n_{jC}) \times t_{ij}}{\sqrt{n_{jE} \times n_{jC}} \times \sqrt{df_{ij}}} \quad (4)$$

Therefore, in our case, instead of directly calculating *Hedges' g*, we first apply *limma* to obtain the moderated t-test and their degrees of freedom. Subsequently, the Equation (4) is used to obtain *Cohen's d*. With *Cohen's d*, we then calculate *Hedges' g* and its variance. This effect size is comparable to the one proposed by *Marot et al.*[24], with the difference that, in this case, the degrees of freedom adjusted by *limma* are considered, whereas in their study, *Cohen's d* is calculated solely from the *moderated Student's t-statistic* and the corresponding sample sizes.

### 2.2.2. Bias correction of the Hedges' g variance

Some authors have noted that the use of the *Hedges' g* (similar to *Cohen's d*) can lead to a bias that tends to inaccurately estimate the combined variance[25,26]. In standard meta-analysis, where only one effect size from various studies is combined, such biases may not significantly impact the results. However, in the context of gene set enrichment meta-analysis, where distinct pathways are evaluated for each study, the bias in the variance of the combined effect can potentially contribute to elevated rates of both, false positives and false negatives, in identified significant non-significant pathways respectively. This is attributed to the fact that minor effect sizes result in small variances (Equation (3)) that can yield significant pathways, while substantial differences lead to larger variances and, consequently, a potential lack of statistical significance. To correct bias in the variance of *Hedges' g*, Lin et al. described an alternative calculation for this based on the mean of the different *Hedges' g* estimators[25] which is based in:

$$V_{g_{ij}}(\bar{g}_i) = \frac{1}{n_{jE}} + \frac{1}{n_{jC}} + \frac{\bar{g}_i^2}{2(n_{jE}+n_{jC})} \quad (5)$$

Where $\bar{g}_i$ is the mean of the different effects sizes for the pathway $i$. Although, as they described, there are other alternative calculations to control this variance's bias[27], the remaining options produce similar results to the Equation (5)[25]. Consequently, we opted to employ this last formula for estimating *Hedges 'g* variance.

### 2.2.3. Random Effects Model for combining effects size

Models based on the combination of these effects sizes aim to obtain a common effect (called combined effect) for all studies[19]. To combine the effects of these studies, two models are distinguished: Fixed Effects Model (FEM) and the Random Effects Model (REM). In our case, we will consider the combined effect obtained from these models as the *combined enrichment score* (CES) value for each of the pathways under study.

FEM is a linear model that assumes the different studies share a common true effect size. The combined effect size is calculated as a weighted mean of the different effect sizes[19]. Therefore, the *combined enrichment score* value for a pathway using this model would be calculated as:

$$CES_i = \frac{\sum_{j=1}^{K} \omega_{ij} Y_{ij}}{\sum_{j=1}^{K} \omega_{ij}} \quad (6)$$

Where:

- $Y_{ij}$ is the pathway effect size of each study. In this particular case, the effects of each pathway in each study are the different *Hedges' g* calculated $(Y_{ij} = g_{ij})$
- $\omega_{ij} = \frac{1}{V_{Y_{ij}}}$ are the different weights assigned to each pathway in each study. $V_{Y_{ij}}$ is the inverse within-study variance, that is, the different $V_{g_{ij}}(\bar{g}_i)$ calculated for each study.

The variance of this combined effect is calculated as:

$$V_{CES_i} = \frac{1}{\sum_{j=1}^{K} \omega_{ij}} \quad (7)$$

The *combined enrichment score* value follows a standard normal, $N(0,1)$:

$$Z_i = \frac{CES_i}{\sqrt{V_{CES_i}}} \quad (8)$$

Therefore, we can obtain a two-tailed p-value:

$$p_i = 2[1 - \Phi|Z_i|] \quad (9)$$

Where $\Phi$ is the standard normal cumulative distribution function.

Unlike FEM, the random-effects model (REM) assumes that the true effect can vary from one study to another. In this case, the combined effect size represents the average of the true effects. In practice, this implies assuming that in the calculation of the weights for the weighted mean, there are two sources of error: the within-study variance (similar to FEM) and the between-study variance ($\tau^2$). To calculate $\tau^2$, we used the method of moments (DerSimonian and Laird)[28]:

$$\tau_i^2 = max\left(0, \frac{Q_i - df}{C_i}\right) \quad (10)$$

Where:

- $Q_i = \sum_{j=1}^{K} \omega_{ij} \times (Y_{ij} - CES_i)^2$, is the total variance. $\omega_i$ are the weights used in the FEM model and $CES_i$ the combined effect obtained in the FEM model.

- $df = K - 1$, is the degrees of freedom, where $K$ is the number of studies.
- $C_i = \sum_{j=1}^{K} \omega_{ij} - \frac{\sum_{j=1}^{K} \omega_{ij}^2}{\sum_{j=1}^{K} \omega_{ij}}$.

Resulting in the weights:

$$\omega_{ij}^* = \frac{1}{V_{Y_{ij}} + \tau_i^2} \qquad (11)$$

Therefore, similarly to the FEM, the combined enrichment score for the REM is calculated as:

$$CES_i^* = \frac{\sum_{j=1}^{K} \omega_{ij}^* g_{ij}}{\sum_{j=1}^{K} \omega_{ij}^*} \qquad (11)$$

FEM should only be used when the studies included in the analysis are functionally identical (not independently conducted) and the results cannot be generalized to other studies[19]. GSA´s studies are not expected to meet this constraint and, therefore, we have decided to focus on the REM model.

## 2.3. Filtering out lowly expressed pathway

Despite our use of the moderated t-test, which significantly reduces bias in the variance of the Hedges' g estimator, false positives may still be founded. These are often attributable to small variances within experimental and control groups associated with pathways exhibiting low activity. To mitigate this issue, pre-filtering pathways with low activity in both control and case groups is a critical step. Pre-filtering is a widely used practice in differential expression analysis, particularly in RNA-Seq data analysis, where genes with very low counts can spuriously appear significantly differentially expressed[29]. The threshold for determining low activity varies depending on the study context and the specific single-sample enrichment (SSE) technique employed. For instance, techniques like GSVA, ssGSEA, and singscore yield scores within a range of -1 to 1, where a value of -1 represents maximum theoretical under-regulation of a gene set and a value of 1 represents maximum over-regulation. However, due to transformations inherent to ssGSEA and singscore, their score distributions are typically more constrained. In contrast, scores derived from the Zscore method follow a standard normal distribution, N(0,1). To establish a common filtering criterion across all SSE techniques, we normalize pathway activity scores when using GSVA, ssGSEA, or singscore. This involves study-specific standardization, where each gene set matrix is normalized by subtracting the mean and dividing by the standard deviation. This ensures that scores across techniques are aligned within a comparable range, simplifying the filtering process. For Zscore-based analyses, this normalization step is unnecessary, as the z-scores are already standardized.

In our study, pathways with an absolute median activity below 0.65 in both control and experimental groups were excluded. While the choice of this threshold may depend on the specific characteristics of a dataset, our selection is justified as follows: in a normal distribution, approximately 50% of the central values fall between -0.65 and 0.65. Filtering out pathways with median activity below this range ensures that we remove those with low expression in both groups, effectively excluding gene sets with minimal activity variation. This approach prioritizes the inclusion of pathways with meaningful activity differences while reducing noise and false positives in the analysis.

## 2.4. Analysis of Simulated Data, SLE and PD gene expression datasets

To evaluate the performance of GSEMA, we tested it alongside other pathway enrichment meta-analysis methods using simulated data and real cases. Simulated bulk RNA-Seq datasets were generated using the Bioconductor package MOsim[30]. Five independent studies were generated, each consisting of 50 control and 50 case samples, and encompassing 39,359 genes. We assumed that 1% of the genes in each study were differentially expressed. To further challenge the methods, 23 differentially overexpressed genes were assigned to a fictitious gene set ("*Simulated_Pathway*").

To test the method with real data, we applied it to three different methods. The first analysis consists in four gene expression datasets of Systemic Erythematosus Lupus (SLE) that are stored in NCBI-GEO with the identifiers GSE108497, GSE61635, GSE65391 and GSE72509. The second study of SLE is composed by five different datasets with identifiers GSE11909_GPL96, GSE11909_GPL97, GSE24706, GSE50772, and GSE82221_GPL10558 from different microarray platforms and a dataset GSE122459 from RNA-Seq.
All datasets, except GSE122459, were directly downloaded from ADEX[31], a database that contains normalized and standardized gene expression data from autoimmune diseases studies.
GSE122459 was obtained from the recount3 database[32] and raw data was preprocessed with the edgeR[33,34] package to filter genes with very low expression and subsequently with trimmed Mean of M values (TMM) from the NOISeq package[35] for normalization.
Finally, to test the utility of GSEMA, it was applied to perform a meta-analysis in PD containing 4 GEO studies from different sequencing platforms. The studies are identified as: GSE6613, GSE18838, GSE22491, GSE54536.

In both simulated and real data, the following methods were applied:

Meta-Analysis and Gene Set Enrichment (MA_GSA): This is the traditional approach. Performing first gene expression meta-analysis at a gene level based on effect size and the resulting gene list is analyzed to find enriched pathways with a Gene Set Analysis (GSA).

Meta-Analysis for Pathway Enrichment (MAPE) Methods[10]:

- MAPE-G: Conducts a differential expression meta-analysis followed by meta-analysis of p-values using Wilkinson's method (also known as the maximum p-value method)[36]. A subsequent GSA is then performed on the integrated results, which is known by other authors as *intermediate merging*[37].
- MAPE-P: Applies differential expression analysis within each study and then performs GSA individually. Finally, these GSA results are combined across studies using Wilkinson's method.
- MAPE-I: Integrates MAPE-P and MAPE-G outcomes by merging p-values using the minimum of p-values method (Tippet's method)[4,10].

GSEMA Methodology: We applied GSEMA using four single-sample enrichment (SSE) methods—GSVA, Z-score, ssGSEA, and singscore—for generating the gene set matrices. For enrichment analyses in MA_GSA and the MAPE methods, we employed the R package fgsea[38]. with the entire MsigDB gene set database, specifically the human canonical gene sets[39], filtering out from the analysis gene sets with less than seven genes. A gene set was defined as significant if its adjusted p-value was < 0.05. Significant sets identified by GSEMA methods were ranked according to the absolute value of their combined effect sizes, while significant sets from MA_GSA and related approaches were ranked by their normalized enrichment scores.

# 3. Results

## 3.1. Gene set enrichment meta-analysis workflow

The proposed approach involves a systematic series of steps to identify significant gene sets (see Figure 1) from diverse studies, each characterized by an expression matrix and a vector that classifies samples into two groups (e.g., experimental and control).

Gene Set Scoring: We apply single-sample enrichment scoring methods—such as ssGSEA, GSVA, Z-score, or singscore—to each independent study. This step produces a set of enrichment matrices, with each matrix containing pathway-specific enrichment scores for every sample. To ensure comparability across studies, these gene set matrices are normalized so that their score ranges align. We then filter out pathways with consistently low expression in both experimental and control groups, as described in Section 2.3.

Effect Size Calculation: For each retained pathway in each study, we use the R package limma to estimate moderated t-statistics and degrees of freedom. These values are then converted into standardized mean differences and their corresponding variances (Section 2.2.1). To further refine these effect size estimates, we apply Equation (5) to correct for potential bias in the variance of the standardized mean differences.

Meta-Analysis: Finally, we integrate effect sizes across all studies using a random-effects meta-analysis model. The resulting p-values are adjusted for multiple testing using the Benjamini–Hochberg method, ensuring a robust control of the false discovery rate[40].

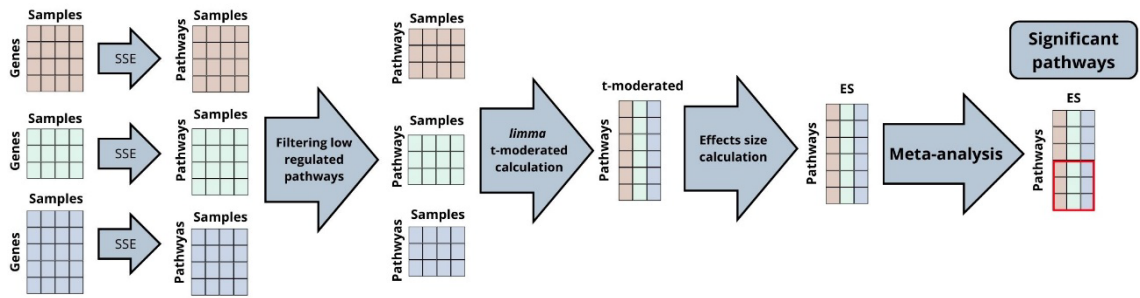

**Figure 1: GSEMA workflow**. The GSEMA workflow begins by applying single-sample enrichment (SEE) techniques to obtain pathway matrices from expression data. Subsequently, pathways with low scores are filtered out and differential expression is computed for each pathway in each study and effect sizes based on Hedges' g are calculated. Finally, a meta-analysis is applied to identify significantly deregulated pathways.

## 3.2. Performance Evaluation Using Simulated Data

To evaluate the performance of our proposed methodology (GSEMA), we compared it with other pathway enrichment meta-analysis methods (described in the Methods section) using simulated data. This data comprised five independent studies, each with 23 significantly differentially expressed genes annotated to a common gene set, termed "Simulated_Pathway," which served as a positive control. Results are summarized in Table 1. All tested methods identified "Simulated_Pathway" as the most influential pathway in the data. Notably, GSEMA, when utilizing ssGSEA and singscore as single-sample enrichment (SSE) methods, also detected a few additional significant gene sets, though these were minimal relative to the total number tested. In contrast,

| Methodology | Number of significant genes set | *"Simulated_Pathway"* top position |
|---|---|---|
| MA_GSA | 1 | 1 |
| MAPE-G | 1 | 1 |
| MAPE-P | 1 | 1 |
| MAPE-I | 1 | 1 |
| GSEMA-GSVA | 1 | 1 |
| GSEMA-Zscore | 1 | 1 |
| GSEMA-ssGSEA | 8 | 1 |
| GSEMA-singscore | 10 | 1 |

**Table 1. Results of applying different methodologies to the simulated data.** The table shows the results of applying various pathway enrichment meta-analysis methodologies to the simulated data. The first column displays the number of significant pathways identified by each method. The second column shows the rank of the Simulated Pathway in each method after ordering all pathways based on the combined effect (GSEMA) or the Normalized Enrichment Score (MAPE, and MA_GSA methods) in absolute value.

To further evaluate the consistency and robustness of the methods, we assessed their false positive rates, which is a well-known limitation of GSEA based methods, using randomly permuted sample classes. Specifically, samples in the simulated data were randomly reassigned to case or control groups, and the frequency of significant pathways was recorded over 100 iterations. Figure 2A illustrates the number of significant gene sets (p-value < 0.05) identified by each method across these simulations

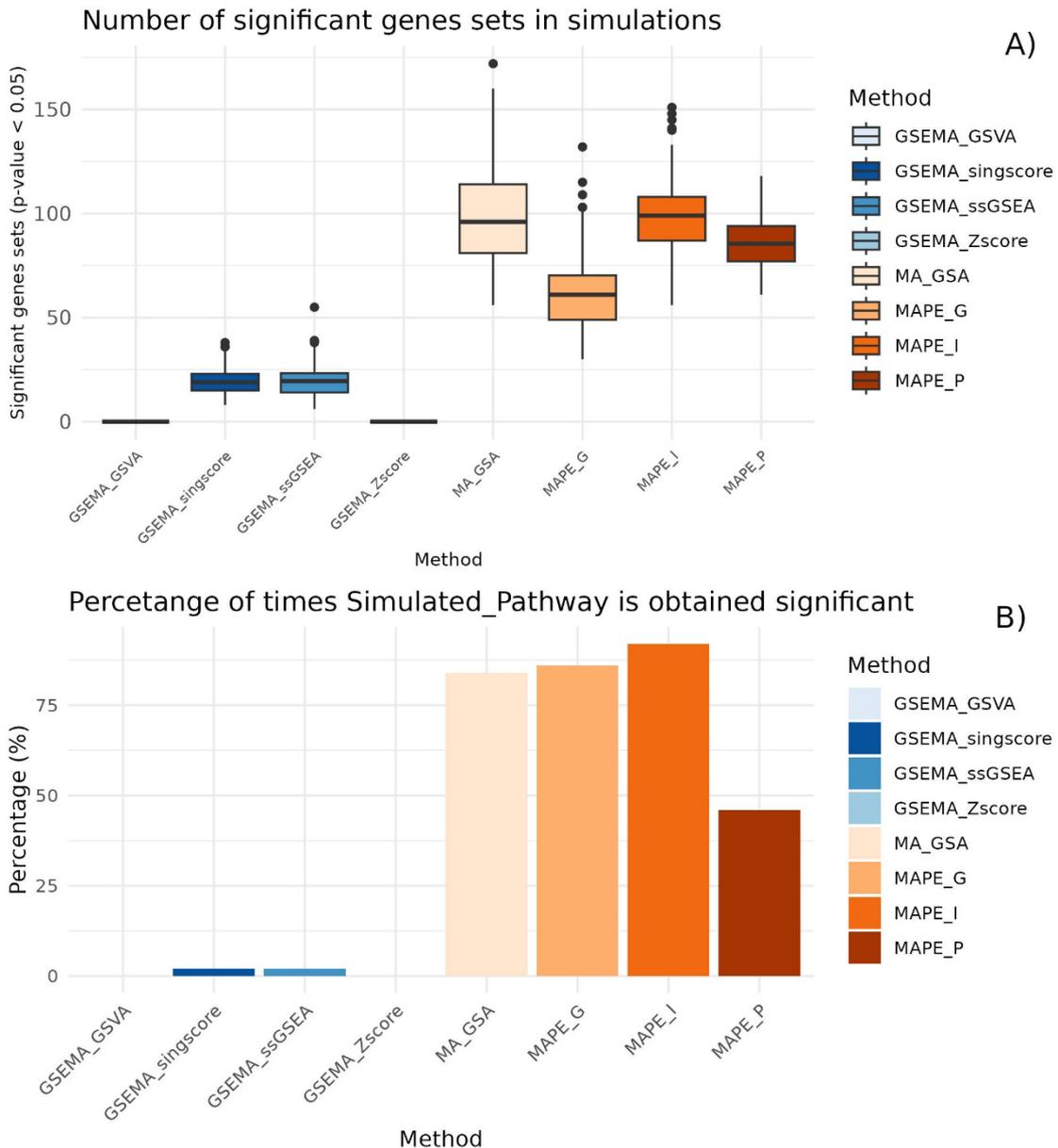

**Figure 2:** A) Boxplot of the number of significant pathways obtained in simulations by each method. For all methods, we conducted 100 simulations by swapping the labels of cases and controls. The figure shows a boxplot of the number of pathways with a p-value < 0.05 obtained in the simulations for each method. B) Bar plot of the percentage of times the "*Simulated_Pathway*" gene set is obtained significant (p-value <0.05) in simulations. For all methods, we conducted 100 simulations by changing the labels of cases and controls.

The permutation analysis revealed that MAPE and MA_GSA also produced over 50 false positives in most cases. By contrast, the GSEMA methods yielded far fewer random significant pathways, suggesting better control over Type I error (false positive rate). In addition, for the specific case of the "Simulated_Pathway," the percentage of times it was identified as significant was much lower for GSEMA (0%–5%) compared to MAPE and MA_GSA (over 45%) (See Figure 2B). These results underscore the effectiveness of GSEMA in controlling false positive rates compared to the other approaches. This trend of performance in controlling false positives and false negatives was also tested using real datasets (see below).

### 3.3. Application to real data

To highlight the advantages of GSEMA, we tested its performance on a range of real-world datasets. Specifically, we applied our methodology to the analysis of Systemic Lupus Erythematosus and Parkinson's disease datasets to confirm the robustness and statistical power of the technique and to illustrate how this methodology can preserve significant information when substantial proportions of missing genes are present.

### 3.3.1. Meta-Analysis of SLE data

To check the power and robustness of GSEMA, we first analyzed four gene expression datasets of Systemic Lupus Erythematosus (SLE). These studies were chosen because their samples derived from the same tissue, which ensures a higher degree of homogeneity among datasets compared to extracting them from different tissues. A summary of the different characteristics of the studies can be found in Table 2.

| Dataset | Healthy samples | Disease samples | Total samples | Tissue |
|---|---|---|---|---|
| GSE108497 | 187 | 325 | 512 | Whole blood |
| GSE61635 | 30 | 79 | 109 | Whole blood |
| GSE65391 | 45 | 116 | 161 | Whole blood |
| GSE72509 | 18 | 99 | 117 | Whole blood |

**Table 2. Description of the studies included.** Characteristics of each of the SLE studies included in the analysis.

In this scenario, we compared samples SLE patients with control samples. The same methodologies were employed as in the case of simulated data (see section 3.2). To observe the reliability of the results from the different methodologies, firstly we examine the number of significant gene sets (FDR<0.05) obtained by each method (Figure 3). In this case, there are no major differences between the GSEMA methods and MA_GSA and MAPE methods. GSEMA_ssGSEA and GSEMA_singscore obtain a similar number of significant gene sets compared to MAPE_I and MAPE_P. On the other hand, GSEMA_Zscore obtains a similar (though slightly lower) number compared to the MA_GSA and MAPE_G methods. Lastly, the GSMA_GSVA method obtains a much lower number of significant gene sets compared to the others. This seems to indicate that the GSEMA_GSVA method may be much more restrictive than the other methods.

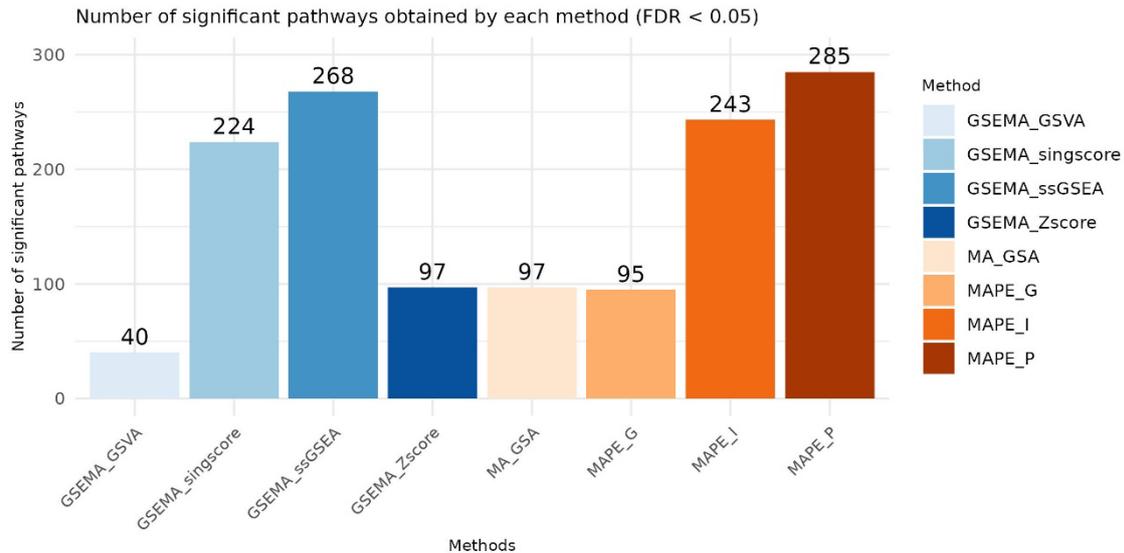

**Figure 3: Bar plot of the number of significant pathways**. A bar plot of the number of significant pathways obtained by each of the pathway enrichment meta-analysis (FDR<0.05).

However, more than the number of significant pathways obtained, what we need to seek is how many of the relevant pathways in those datasets are detected by the methods, and the importance attributed between all the non-relevant gene sets. In the case of SLE, it is well known that genes related with pathways involved in immune response, and specifically, the interferon signature, are overexpressed in patients with active SLE disease. We rank the significant pathways based on the combined effect (GSEMA methods) or the NES (Normalized Enrichment Score, for both MAPE and MA_GSA methods). To simplify the results, we show the top 5 overexpressed pathways obtained by each of the methods (Table 3). In all methods, both GSEMA and MA_GSA, as well as MAPE, gene sets related to immune response, or the interferon signature are obtained. Furthermore, in all of them, the gene set "Reactome Interferon Alpha Beta Signaling" is obtained as the most relevant pathway. In this specific case, the only differences are observed in the databases to which the most relevant pathways belong, with more variability in the GSEMA methods and REACTOME being practically the only one in the rest of the methods.

| Method | Top dysregulated pathways | Related with immune system |
|---|---|---|
| MA_GSA | *REACTOME_INTERFERON_ALPHA_BETA_SIGNALING* | Yes |
|  | *REACTOME_INTERFERON_SIGNALING* | Yes |
|  | *REACTOME_INTERFERON_GAMMA_SIGNALING* | Yes |
|  | *REACTOME_NEUTROPHIL_DEGRANULATION* | Yes |
|  | *REACTOME_CYTOKINE_SIGNALING_IN_IMMUNE_SYSTEM* | Yes |
| MAPE_G | *REACTOME_INTERFERON_ALPHA_BETA_SIGNALING* | Yes |
|  | *REACTOME_INTERFERON_SIGNALING* | Yes |
|  | *REACTOME_INTERFERON_GAMMA_SIGNALING* | Yes |
|  | *REACTOME_NEUTROPHIL_DEGRANULATION* | Yes |
|  | *REACTOME_CYTOKINE_SIGNALING_IN_IMMUNE_SYSTEM* | Yes |
| MAPE-P | *REACTOME_INTERFERON_ALPHA_BETA_SIGNALING* | Yes |
|  | *REACTOME_INTERFERON_SIGNALING* | Yes |
|  | *WP_TYPE_II_INTERFERON_SIGNALING* | Yes |
|  | *WP_TYPE_1_INTERFERON_INDUCTION_AND_SIGNALING_DURING_SARS_COV_2_INFECTION* | Yes |
|  | *REACTOME_INTERFERON_GAMMA_SIGNALING* | Yes |
| MAPE_I | *REACTOME_INTERFERON_ALPHA_BETA_SIGNALING* | Yes |
|  | *REACTOME_INTERFERON_SIGNALING* | Yes |
|  | *REACTOME_INTERFERON_GAMMA_SIGNALING* | Yes |
|  | *REACTOME_NEUTROPHIL_DEGRANULATION* | Yes |
| | REACTOME_CYTOKINE_SIGNALING_IN_IMMUNE_SYSTEM | Yes |
| | WP_IMMUNE_RESPONSE_TO_TUBERCULOSIS | Yes |
| GSEMA_ GSVA | REACTOME_INTERFERON_ALPHA_BETA_SIGNALING | Yes |
| | WP_TYPE_II_INTERFERON_SIGNALING | Yes |
| | WP_IMMUNE_RESPONSE_TO_TUBERCULOSIS | Yes |
| | WP_TYPE_I_INTERFERON_INDUCTION_AND_SIGNALING_DURING_SARS_COV_2_INFECTION | Yes |
| | WP_HOST_PATHOGEN_INTERACTION_OF_HUMAN_CORONAVIRUSES_INTERFERON_INDUCTION | Yes |
| GSEMA_ssGSEA | REACTOME_INTERFERON_ALPHA_BETA_SIGNALING | Yes |
| | WP_TYPE_I_INTERFERON_INDUCTION_AND_SIGNALING_DURING_SARS_COV_2_INFECTION | Yes |
| | WP_HOST_PATHOGEN_INTERACTION_OF_HUMAN_CORONAVIRUSES_INTERFERON_INDUCTION | Yes |
| | REACTOME_OAS_ANTIVIRAL_RESPONSE | Yes |
| | WP_IMMUNE_RESPONSE_TO_TUBERCULOSIS | Yes |
| GSEMA_ Zscore | REACTOME_INTERFERON_ALPHA_BETA_SIGNALING | Yes |
| | WP_IMMUNE_RESPONSE_TO_TUBERCULOSIS | Yes |
| | WP_TYPE_II_INTERFERON_SIGNALING | Yes |
| | WP_TYPE_I_INTERFERON_INDUCTION_AND_SIGNALING_DURING_SARS_COV_2_INFECTION | Yes |
| | KEGG_MEDICUS_REFERENCE_TYPE_I_IFN_SIGNALING_PATHWAY | Yes |
| GSEMA_singscore | REACTOME_INTERFERON_ALPHA_BETA_SIGNALING | Yes |
| | REACTOME_OAS_ANTIVIRAL_RESPONSE | Yes |
| | WP_IMMUNE_RESPONSE_TO_TUBERCULOSIS | Yes |
| | WP_HOST_PATHOGEN_INTERACTION_OF_HUMAN_CORONAVIRUSES_INTERFERON_INDUCTION | Yes |
| | WP_TYPE_I_INTERFERON_INDUCTION_AND_SIGNALING_DURING_SARS_COV_2_INFECTION | Yes |

**Table 3. The most significant pathways obtained by each of the enrichment pathway meta-analysis methods for SLE data.** The table shows the five most enriched pathways obtained by each of the methods. The table includes the five pathways with the highest absolute values of CES or NES obtained by each of the methods. It also indicates whether the identified pathways are related to the immune system.

To contrast the robustness and consistency of the results, we conducted a simulation analysis similar to the one used in simulated data. In this case, 100 simulations were conducted (100 in the case of ssGSEA and GSVA methods) in which the labels of the patients' conditions (SLE and healthy) were randomly swapped. In the same way as in the case of simulated data, the number of significant pathways (with a p-value less than 0.05) obtained by each of the methods was observed (Figure 4). In Figure 4, it can be observed that the GSEMA methods yield remarkably fewer significant pathways in the simulations, demonstrating much more robust results. The only drawback is that in the case of GSEMA_GSVA, no significant pathways are obtained, which may indicate a power issue.

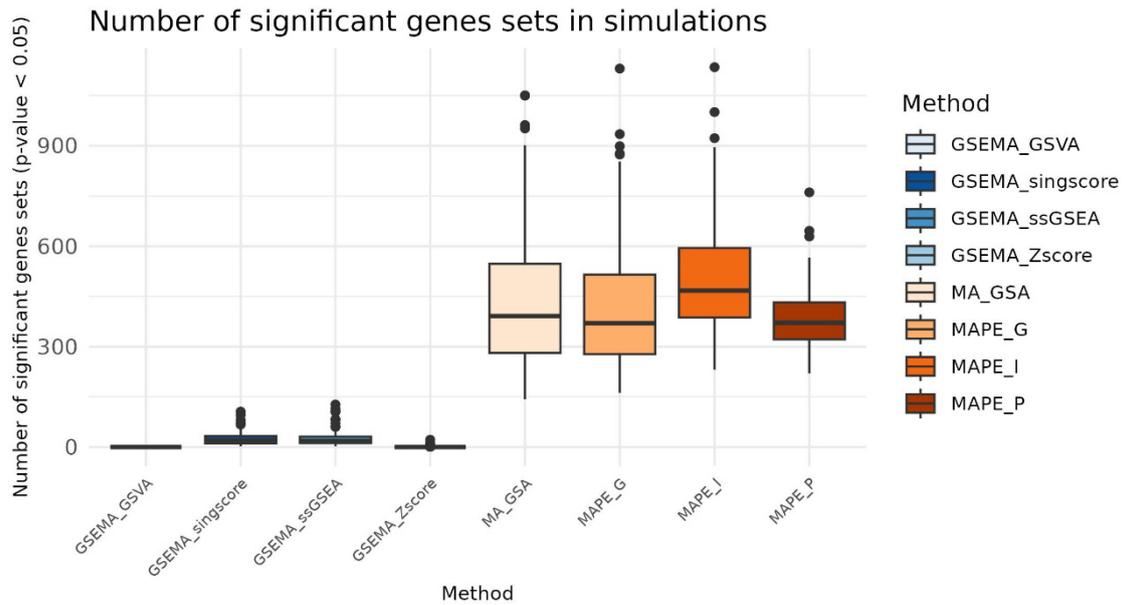

**Figure 4**: **Boxplot of the number of significant pathways obtained in simulations by each method in SLE data**. For all methods, we conducted 100 simulations by swapping the labels of cases and controls. The figure shows a boxplot of the number of pathways with a p-value < 0.05 obtained in the simulations for each method.

As can be seen in the results obtained, for almost all GSEMA methods, the number of significant gene sets is 0 in most permutations, whereas this number varies considerably for the MAPE methods. This confirms that the results obtained by GSEMA are much more consistent. This also allows us to identify that the presence of false positives in GSEMA may be more due to the SSE methods themselves than to the GSEMA methodology, implying that better sample enrichment techniques could improve GSEMA´s performance.

### 3.3.2. Performance of GSEMA to combine studies with missing genes

One of the key advantages of GSEMA is its ability to preserve biological information even when some genes are missing in one or more datasets. Additionally, it facilitates the integration of microarray and RNA-Seq studies by promoting greater data harmonization using single-sample enrichment techniques. To evaluate this, we analyzed datasets with missing genes or non-matching gene lists across studies. Specifically, we selected six SLE datasets (see Table 4). Some of these studies were considered in other previous meta-analyses[9,41], although never simultaneously.

| Dataset | Healthy samples | Disease samples | Tissue (Cell type) | Platform | Number of genes |
|---|---|---|---|---|---|
| GSE11909_GPL96 | 10 | 118 | Peripheral blood (PBMCs) | GPL96 (microarray) | 13882 |
| GSE11909_GPL97 | 7 | 53 | Peripheral blood (PBMCs) | GPL97 (microarray) | 11234 |
| GSE24706 | 33 | 15 | Peripheral blood (PBMCs) | GPL6884 (microarray) | 12467 |
| GSE50772 | 20 | 61 | Peripheral blood (PBMCs) | GPL570 (microarray) | 21767 |
| GSE82221_GPL10558 | 25 | 30 | Peripheral blood (PBMCs) | GPL10558 (microarray) | 14419 |

| GSE122459 | 6 | 20 | Peripheral blood (PBMCs) | GPL16791 (RNA-Seq) | 24123 |

**Table 4. Description of the studies included**. Characteristics of each of the SLE studies of different platforms with missing genes included in the analysis.

As mentioned earlier, typically, when conducting a gene expression meta-analysis of these datasets, we usually work with the common genes. In this case, the number of common genes across these studies is 3342. This can bias the results, as can be observed when applying a gene expression meta-analysis based on effect size combination followed by a GSA (MA_GSA), where we do not observe any gene set associated with either the interferon signature or immune response in the top 5 positions (Table 5). A similar result is observed when applying the MAPE_G and MAPE_I methods to this same dataset. Only the MAPE_P method yields results with pathways related to immune and interferon responses as some of the most important ones. Nevertheless, when we perform the different GSEMA techniques, gene sets related to interferon and immune response are obtained among the most important ones. Only GSEMA, along with GSVA as the SSE technique, yields few gene sets related to immune response and interferon. This corroborates the good applicability that GSEMA can have in this type of meta-analysis as well as when combining RNA-Seq and microarray datasets. Additionally, the fact that MAPE_P obtains certain relevant results also demonstrates the importance of considering missing genes, as this technique, by performing a GSA in each dataset, never works directly with common genes.

| Method | Top dysregulated pathways | Related with immune system |
|---|---|---|
| MA_GSA | *REACTOME_EUKARYOTIC_TRANSLATION_ELONGATION* | No |
| | *KEGG_MEDICUS_REFERENCE_TRANSLATION_INITIATION* | No |
| | *WP_CYTOPLASMIC_RIBOSOMAL_PROTEINS* | No |
| | *REACTOME_EUKARYOTIC_TRANSLATION_INITIATION* | No |
| | *REACTOME_RESPONSE_OF_EIF2AK4_GCN2_TO_AMINO_ACID_DEFICIENCY* | No |
| MAPE_G | *KEGG_MEDICUS_REFERENCE_TRANSLATION_INITIATION* | No |
| | *WP_CYTOPLASMIC_RIBOSOMAL_PROTEINS* | No |
| | *REACTOME_EUKARYOTIC_TRANSLATION_INITIATION* | No |
| | *REACTOME_RESPONSE_OF_EIF2AK4_GCN2_TO_AMINO_ACID_DEFICIENCY* | No |
| | *KEGG_MEDICUS_REFERENCE_TRANSLATION_INITIATION* | No |
| MAPE-P | *REACTOME_INTERFERON_ALPHA_BETA_SIGNALING* | Yes |
| | *REACTOME_NEUTROPHIL_DEGRANULATION* | Yes |
| | *KEGG_MEDICUS_REFERENCE_TRANSLATION_INITIATION* | No |
| | *REACTOME_EUKARYOTIC_TRANSLATION_ELONGATION* | No |
| | *WP_CYTOPLASMIC_RIBOSOMAL_PROTEINS* | No |
| GSEMA_GSVA | *WP_TYPE_II_INTERFERON_SIGNALING* | Yes |
| | *WP_MYD88_DISTINCT_INPUT_OUTPUT_PATHWAY* | No |
| | *KEGG_MEDICUS_REFERENCE_ORGANIZATION_OF_THE_OUTER_KINETOCHORE* | No |
| | *WP_PHOTODYNAMIC_THERAPY_INDUCED_HIF_1_SURVIVAL_SIGNALING* | No |
| | | |
| GSEMA_ssGSEA | *REACTOME_INTERFERON_ALPHA_BETA_SIGNALING* | Yes |
| | *WP_TYPE_I_INTERFERON_INDUCTION_AND_SIGNALING_DURING_SARS_COV_2_INFECTION* | Yes |
| | *WP_TYPE_II_INTERFERON_SIGNALING* | Yes |
| | *WP_HOST_PATHOGEN_INTERACTION_OF_HUMAN_CORONAVIRUSES_INTERFERON_INDUCTION* | Yes |

| | WP_IMMUNE_RESPONSE_TO_TUBERCULOSIS | Yes |
|---|---|---|
| **GSEMA_Zscore** | WP_TYPE_I_INTERFERON_INDUCTION_AND_SIGNALING_DURING_SARS_COV_2_INFECTION | Yes |
| | WP_TYPE_II_INTERFERON_SIGNALING | Yes |
| | WP_NETWORK_MAP_OF_SARS_COV_2_SIGNALING_PATHWAY | No |
| | REACTOME_INTERFERON_ALPHA_BETA_SIGNALING | Yes |
| | WP_IMMUNE_RESPONSE_TO_TUBERCULOSIS | Yes |
| **GSEMA_singscore** | REACTOME_INTERFERON_ALPHA_BETA_SIGNALING | Yes |
| | WP_TYPE_I_INTERFERON_INDUCTION_AND_SIGNALING_DURING_SARS_COV_2_INFECTION | Yes |
| | WP_TYPE_II_INTERFERON_SIGNALING | Yes |
| | WP_HOST_PATHOGEN_INTERACTION_OF_HUMAN_CORONAVIRUSES_INTERFERON_INDUCTION | Yes |
| | WP_IMMUNE_RESPONSE_TO_TUBERCULOSIS | Yes |

**Table 5:** Summary of the most important gene sets obtained by each method. The table includes the five pathways with the highest absolute values of CES or NES obtained by each of the methods. It also indicates whether the identified pathways are related to the immune system (Green) or not (red).

### 3.3.3. Identification of novel deregulated pathways in the analysis of Parkinson's blood samples

Finally, to test the utility of GSEMA, it was applied to perform a meta-analysis carried out by *Santiago and Potashkin*[42] in which 4 GEO studies from different sequencing platforms were combined (Table 6). In this study, different biomarkers of the disease were identified, and various upregulated pathways were associated. Specifically, the differentially up-regulated pathways obtained were: bacterial invasion of epithelial cells, mitogen-activated protein kinase-signaling pathway, fructose and mannose metabolism, T-cell receptor-signaling pathway, mammalian target of rapamycin-signaling pathway, type 2 diabetes mellitus, and colorectal cancer.

| Dataset | Healthy samples | Disease samples | Platform | Tissue |
|---|---|---|---|---|
| GSE6613 | 50 | 22 | Affymetrix Human Genome U133A | Peripheral blood |
| GSE18838 | 18 | 12 | Affymetrix Exon Arrays | Peripheral blood |
| GSE22491 | 10 | 7 | Agilent Whole Human Genome | Peripheral blood |
| GSE54536 | 5 | 5 | Ilumina HT-12 V4 | Peripheral blood |

**Table 6. Description of the studies included**. Characteristics of each of the Parkison's disease studies of different platforms with missing genes included in the analysis.

In our case, we replicated the analysis using GSEMA and focused on the pathways that were also found to be up-regulated. In this case, ssGSEA was applied as a single-sample enrichment technique, and no filtering of the pathways was carried out since only the pathways available in the KEGG database were considered. The up-regulated pathways obtained through GSEMA can be seen in Table 7.

| Pathway | Combined Enrichment Score |
|---|---|
| Neuroactive ligand-receptor interaction | 0.8608 |
| Steroid hormone biosynthesis | 0.6036 |

| Insulin secretion | 0.4440 |
| Thyroid hormone synthesis | 0.4397 |
| Hypertrophic cardiomyopathy | 0.4146 |
| Circadian entrainment | 0.3938 |
| Calcium signaling pathway | 0.3682 |

**Table 7. Significant up-regulated pathways identified through the application of the GSEMA method with ssGSEA SSE on Parkinson's datasets**. This table presents the significant pathways obtained from the pathway enrichment meta-analysis using GSEMA with the ssGSEA SSE method, along with the corresponding Combined Effect Sizes (CES)

GSEMA has identified differentially up-regulated pathways that may be related to Parkinson's disease. For example, the *Neuroactive ligand-receptor interaction* pathway is the one with the highest CES and refers to a set of molecular interactions between neuroactive ligands and their corresponding receptors in the nervous system. In the literature, we can find that the dysregulation of this pathway is associated with neurodegenerative diseases such as Parkinson's. It may be notable that this pathway was not identified in the study by *Santiago and Potashkin*. To investigate why this pathway was not detected, we conducted a meta-analysis considering genes present in at least two of the studies using the DExMA[9] package. In this meta-analysis, we identified that, out of the 367 genes that make up the pathway, 37 significant genes were obtained. However, in the common genes approach (the one conducted in the article), 8 genes (*GPR156, KISS1R, LYNX1, NPBWR1, RXFP4, UCN2, UCN3,* and *UTS2R*) from these 37 were not considered, which could significantly affect the final functional enrichment analysis.

Moreover, other pathways have been identified that the literature shows are altered in Parkinson's disease but were also not found in the article, such as *Steroid hormone biosynthesis*[43], *Thyroid hormone synthesis*[44], *Circadian entrainment*[45] and *Calcium signaling pathway*[46]. This highlights that, when combining studies from different platforms, we may lose information in the final results if we only consider the genes common to all studies. These results not only highlight the potential loss of information in the final results when combining studies from different platforms by considering only the common genes, but also demonstrate that the use of GSEMA better preserves biological information and allows for the identification of significantly dysregulated pathways related to the condition under study.

## 4. Discussion

In recent years, the growing number of datasets available in public repositories has amplified the importance of meta-analysis techniques in transcriptomics research. These methods have become increasingly popular for integrating gene expression datasets and uncovering new biological insights. However, applying gene expression meta-analysis to combine data from different transcriptome platforms poses significant challenges. One potential source of error identified when combining this type of data is the presence of missing or unmeasured genes[7]. This issue can be especially evident when integrating microarray and RNA-Seq studies, where data heterogeneity can lead to the loss of biological information and introduce bias.

This lack of data and homogeneity can lead to the loss of biological information and biased results acquisition. One way to address this issue is to shift the focus from individual gene expression to biological information encoded by gene sets. This approach allows for the integration of gene set data rather than individual genes, mitigating the effects of missing data and enhancing robustness.

While enrichment analysis is widely employed in gene expression data analysis, its use in conjunction with meta-analysis techniques remains underexplored. Although several tools have been developed for such analyses, few, apart from iGSEA, utilize meta-analysis methods based on effect size—a standard approach in meta-analysis. Instead, most methods combine p-values from gene set analyses, which often results in the loss of directionality[4], that is, activation or repression patterns. Recent advancements in single-sample enrichment scoring techniques, which compute enrichment scores for individual samples and gene sets, have opened new avenues for pathway enrichment meta-analysis. These techniques enable the identification of critical gene sets across diverse datasets.

In this work, we introduce GSEMA, a novel methodology that integrates effect size-based meta-analysis with single-sample enrichment scoring to perform pathway enrichment meta-analysis across multiple studies. GSEMA addresses the potential biases inherent in effect size-based meta-analysis and accommodates non-matching genes across datasets. Its application to both simulated and real datasets demonstrates its ability to control false positives and identify more specific pathways of interest. Our results indicate that GSEMA provides more robust and consistent results than common pathway enrichment meta-analyses techniques. Moreover, by working with gene sets rather than the direct expression of genes, it is not affected by the possible existence of missing genes in each one of the datasets. This, in turn, allows the combination of studies from different platforms since gene sets are less affected when combining techniques using a comparable enrichment score across studies.

Despite its strengths, GSEMA also has certain limitations. Single-sample enrichment scoring techniques can be computationally intensive for large datasets, making the transformation from expression matrices to gene set matrices time-consuming in some cases. Additionally, the optimal technique for calculating pathway matrices and the appropriate filtering thresholds depend on the specific characteristics of the studies, requiring careful fine-tuning.

GSEMA is a methodology that enables the integration of multiple studies from different platforms and, to the best of our knowledge, is the first to merge single-sample enrichment scoring techniques with meta-analysis methods. Additionally, although it has been applied within the context of gene expression analysis, we would like to emphasize that this methodology can be extended to other -omics fields such as methylation or proteomics.

# Authorship

**JAVG:** Conceptualization; Methodology; Software; Formal Analysis; Writing – original draft; writing – review and editing; formal analysis; data curation. **PPJB:** Methodology; Software; Formal Analysis; Writing – review and editing; formal analysis. **PCS:** Conceptualization; Supervision; Methodology; Writing – review and editing.

# Funding

This work has been funded by grant PID2020-119032RB-I00 funded by MCIN/AEI/10.13039/501100011033. JAVG was funded by the Teaching Staff Program, Ministerio de Universidades (grant number FPU19/01999).

# Conflicts of Interest

The authors declare no conflict of interest.

# Data availability

Gene expression data is publicly available at GEO repository. GSEMA R package is available at CRAN and GitHub repository.

# Acknowledgments

This work is part of the thesis of Pablo Jurado-Bascón in the doctorate program of Mathematical and Applied Statistics of the University of Granada.

# Highlights

**What is already known?**

**What is new?**

**Potential impact for Research Synthesis Methods readers**

# References

1. Perez-Riverol Y, Zorin A, Dass G, et al. Quantifying the impact of public omics data. *Nat Commun*. 2019;10:3512. doi:10.1038/s41467-019-11461-w

2. NCBI Resource Coordinators. Database resources of the National Center for Biotechnology Information. *Nucleic Acids Res*. 2018;46(D1):D8-D13. doi:10.1093/nar/gkx1095

3. Athar A, Füllgrabe A, George N, et al. ArrayExpress update – from bulk to single-cell expression data. *Nucleic Acids Res*. 2019;47(Database issue):D711-D715. doi:10.1093/nar/gky964

4. Toro-Domínguez D, Villatoro-García JA, Martorell-Marugán J, Román-Montoya Y, Alarcón-Riquelme ME, Carmona-Sáez P. A survey of gene expression meta-analysis: methods and applications. *Brief Bioinform*. 2021;22(2):1694-1705. doi:10.1093/bib/bbaa019

5. Subramanian A, Tamayo P, Mootha VK, et al. Gene set enrichment analysis: a knowledge-based approach for interpreting genome-wide expression profiles. *Proc Natl Acad Sci U S A*. 2005;102(43):15545-15550. doi:10.1073/pnas.0506580102

6. Fung WT, Wu JT, Chan WMM, Chan HH, Pang H. Pathway-based meta-analysis for partially paired transcriptomics analysis. *Research Synthesis Methods*. 2020;11(1):123-133. doi:10.1002/jrsm.1381

7. Bobak CA, McDonnell L, Nemesure MD, Lin J, Hill JE. Assessment of Imputation Methods for Missing Gene Expression Data in Meta-Analysis of Distinct Cohorts of Tuberculosis Patients. *Pac Symp Biocomput*. 2020;25:307-318.

8. Mancuso CA, Canfield JL, Singla D, Krishnan A. A flexible, interpretable, and accurate approach for imputing the expression of unmeasured genes. *Nucleic Acids Res*. doi:10.1093/nar/gkaa881

9. Villatoro-García JA, Martorell-Marugán J, Toro-Domínguez D, Román-Montoya Y, Femia P, Carmona-Sáez P. DExMA: An R Package for Performing Gene Expression Meta-Analysis with Missing Genes. *Mathematics*. 2022;10(18):3376. doi:10.3390/math10183376

10. Shen K, Tseng GC. Meta-analysis for pathway enrichment analysis when combining multiple genomic studies. *Bioinformatics*. 2010;26(10):1316-1323. doi:10.1093/bioinformatics/btq148

11. Chen M, Zang M, Wang X, Xiao G. A powerful Bayesian meta-analysis method to integrate multiple gene set enrichment studies. *Bioinformatics*. 2013;29(7):862-869. doi:10.1093/bioinformatics/btt068

12. Lu W, Wang X, Zhan X, Gazdar A. Meta-analysis approaches to combine multiple gene set enrichment studies. *Stat Med*. 2018;37(4):659-672. doi:10.1002/sim.7540

13. Barbie DA, Tamayo P, Boehm JS, et al. Systematic RNA interference reveals that oncogenic KRAS-driven cancers require TBK1. *Nature*. 2009;462(7269):108-112. doi:10.1038/nature08460

14. Hänzelmann S, Castelo R, Guinney J. GSVA: gene set variation analysis for microarray and RNA-Seq data. *BMC Bioinformatics*. 2013;14(1):7. doi:10.1186/1471-2105-14-7

15. Lee E, Chuang HY, Kim JW, Ideker T, Lee D. Inferring Pathway Activity toward Precise Disease Classification. *PLOS Computational Biology*. 2008;4(11):e1000217. doi:10.1371/journal.pcbi.1000217

16. Foroutan M, Bhuva DD, Lyu R, Horan K, Cursons J, Davis MJ. Single sample scoring of molecular phenotypes. *BMC Bioinformatics*. 2018;19(1):404. doi:10.1186/s12859-018-2435-4

17. Demerath NJ. The American Soldier: Volume I, Adjustment During Army Life. By S. A. Stouffer, E. A. Suchman, L. C. DeVinney, S. A. Star, R. M. Williams, Jr. Volume II, Combat and Its Aftermath. By S. A. Stouffer, A. A. Lumsdaine, M. H. Lumsdaine, R. M. Williams, Jr., M. B. Smith, I. L. Janis, S. A. Star, L. S. Cottrell, Jr. Princeton, New Jersey: Princeton University Press, 1949. Vol. I, 599 pp., Vol. II, 675 pp. $7.50 each; $13.50 together. *Social Forces*. 1949;28(1):87-90. doi:10.2307/2572105

18. Heard NA, Rubin-Delanchy P. Choosing between methods of combining $p$-values. *Biometrika*. 2018;105(1):239-246. doi:10.1093/biomet/asx076

19. Borenstein M, Hedges LV, Higgins JPT, Rothstein HR. *Introduction to Meta-Analysis*. John Wiley & Sons; 2021.

20. Smyth GK. Linear models and empirical bayes methods for assessing differential expression in microarray experiments. *Stat Appl Genet Mol Biol*. 2004;3:Article3. doi:10.2202/1544-6115.1027

21. Ritchie ME, Phipson B, Wu D, et al. limma powers differential expression analyses for RNA-sequencing and microarray studies. *Nucleic Acids Res*. 2015;43(7):e47. doi:10.1093/nar/gkv007

22. Smyth G, Hu Y, Ritchie M, et al. limma: Linear Models for Microarray Data. Published online 2021. doi:10.18129/B9.bioc.limma

23. Rosenthal R, Rosnow RL. *Essentials of Behavioral Research: Methods and Data Analysis*. Third Edition. New York: McGraw-Hill; 2008.

24. Marot G, Foulley JL, Mayer CD, Jaffrézic F. Moderated effect size and P-value combinations for microarray meta-analyses. *Bioinformatics*. 2009;25(20):2692-2699. doi:10.1093/bioinformatics/btp444

25. Lin L, Aloe AM. Evaluation of various estimators for standardized mean difference in meta-analysis. *Stat Med*. 2021;40(2):403-426. doi:10.1002/sim.8781

26. Doncaster CP, Spake R. Correction for bias in meta-analysis of little-replicated studies. *Methods in Ecology and Evolution*. 2018;9(3):634-644. doi:10.1111/2041-210X.12927

27. Hedges LV. Fitting Categorical Models to Effect Sizes from a Series of Experiments. *Journal of Educational Statistics*. 1982;7(2):119-137.

28. DerSimonian R, Kacker R. Random-effects model for meta-analysis of clinical trials: an update. *Contemp Clin Trials*. 2007;28(2):105-114. doi:10.1016/j.cct.2006.04.004

29. Conesa A, Madrigal P, Tarazona S, et al. A survey of best practices for RNA-seq data analysis. *Genome Biology*. 2016;17(1):13. doi:10.1186/s13059-016-0881-8

30. Tarazona Campos S, Martínez-Mira C, Conesa A. Mosim: Multi-Omics Simulation in R. Published online January 10, 2023. doi:10.2139/ssrn.4317583

31. Martorell-Marugán J, López-Domínguez R, García-Moreno A, et al. A comprehensive database for integrated analysis of omics data in autoimmune diseases. *BMC Bioinformatics*. 2021;22(1):343. doi:10.1186/s12859-021-04268-4

32. Wilks C, Zheng SC, Chen FY, et al. recount3: summaries and queries for large-scale RNA-seq expression and splicing. *Genome Biol*. 2021;22(1):323. doi:10.1186/s13059-021-02533-6

33. Chen Y, Chen L, Lun ATL, Baldoni PL, Smyth GK. edgeR 4.0: powerful differential analysis of sequencing data with expanded functionality and improved support for small counts and larger datasets. Published online January 24, 2024:2024.01.21.576131. doi:10.1101/2024.01.21.576131

34. Robinson MD, McCarthy DJ, Smyth GK. edgeR: a Bioconductor package for differential expression analysis of digital gene expression data. *Bioinformatics*. 2010;26(1):139-140. doi:10.1093/bioinformatics/btp616

35. Tarazona S, Furió-Tarí P, Turrà D, et al. Data quality aware analysis of differential expression in RNA-seq with NOISeq R/Bioc package. *Nucleic Acids Res*. 2015;43(21):e140. doi:10.1093/nar/gkv711

36. Li J, Tseng GC. An adaptively weighted statistic for detecting differential gene expression when combining multiple transcriptomic studies. *Ann Appl Stat*. 2011;5(2A):994-1019. doi:10.1214/10-AOAS393

37. Kosch R, Jung K. Conducting gene set tests in meta-analyses of transcriptome expression data. *Res Synth Methods*. 2019;10(1):99-112. doi:10.1002/jrsm.1337

38. Korotkevich G, Sukhov V, Budin N, Shpak B, Artyomov MN, Sergushichev A. Fast gene set enrichment analysis. Published online February 1, 2021:060012. doi:10.1101/060012

39. Liberzon A, Birger C, Thorvaldsdóttir H, Ghandi M, Mesirov JP, Tamayo P. The Molecular Signatures Database (MSigDB) hallmark gene set collection. *Cell Syst*. 2015;1(6):417-425. doi:10.1016/j.cels.2015.12.004

40. Benjamini Y, Hochberg Y. Controlling the False Discovery Rate: A Practical and Powerful Approach to Multiple Testing. *Journal of the Royal Statistical Society: Series B (Methodological)*. 1995;57(1):289-300. doi:10.1111/j.2517-6161.1995.tb02031.x

41. Kröger W, Mapiye D, Entfellner JBD, Tiffin N. A meta-analysis of public microarray data identifies gene regulatory pathways deregulated in peripheral blood mononuclear cells from individuals with Systemic Lupus Erythematosus compared to those without. *BMC Med Genomics*. 2016;9:66. doi:10.1186/s12920-016-0227-0

42. Santiago JA, Potashkin JA. Network-based metaanalysis identifies HNF4A and PTBP1 as longitudinally dynamic biomarkers for Parkinson's disease. *Proceedings of the National Academy of Sciences*. 2015;112(7):2257-2262. doi:10.1073/pnas.1423573112

43. Luchetti S, Bossers K, Frajese GV, Swaab DF. Neurosteroid Biosynthetic Pathway Changes in Substantia Nigra and Caudate Nucleus in Parkinson's Disease. *Brain Pathology*. 2010;20(5):945. doi:10.1111/j.1750-3639.2010.00396.x

44. Mohammadi S, Dolatshahi M, Rahmani F. Shedding light on thyroid hormone disorders and Parkinson disease pathology: mechanisms and risk factors. *J Endocrinol Invest*. 2021;44(1):1-13. doi:10.1007/s40618-020-01314-5

45. Hunt J, Coulson EJ, Rajnarayanan R, Oster H, Videnovic A, Rawashdeh O. Sleep and circadian rhythms in Parkinson's disease and preclinical models. *Molecular Neurodegeneration*. 2022;17(1):2. doi:10.1186/s13024-021-00504-w

46. Calì T, Ottolini D, Brini M. Calcium signaling in Parkinson's disease. *Cell Tissue Res*. 2014;357(2):439-454. doi:10.1007/s00441-014-1866-0